# Sexual videos in Internet: a test of 11 hypotheses about intimate practices and gender interactions in Latin America

Julián Monge-Nájera and Karla Vega Corrales

Laboratorio de Ecología Urbana, Vicerrectoría de Investigación, Universidad Estatal a Distancia, 2050 San José, Costa Rica; julianmonge@gmail.com, kantr.vega@gmail.com

**ABSTRACT**

There is a marked lack of literature on user-submitted sexual videos from Latin America. To start filling that gap, we present a formal statistical testing of several hypotheses about the characteristics of 214 videos from Nereliatube.com posted from the inauguration of the site until December 2010. We found that in most cases the video was made consensually and the camera was operated by the man. The most frequent practice shown was fellatio, followed by vaginal penetration. The great majority of videos showed the sexual interactions of one woman with one man; group sex was rare. Violence and manifestations of power were rare and when there was violence it was mostly simulated. Latin American user-submitted sexual videos in Nereliatube generally reflect a society in which women and men have a variety of sexual practices that are mostly consensual and that do not differ from the biologically and anthropologically expected patterns.

**RESUMEN**

Existe muy poca literatura sobre causas biológicas de expresiones culturales relacionadas con la sexualidad en Internet, y aun menos sobre videos sexuales latinoamericanos. Aquí presentamos una evaluación estadística de once hipótesis basada en 214 videos disponibles en 2010 en el sitio Nereliatube. En la mayoría de los videos ambas partes estaban conscientes de la grabación y el hombre operaba la cámara. La práctica más frecuente fue la felación, seguida de penetración vaginal. La mayoría presentaban una mujer y un hombre, siendo escasos los videos de sexo en grupo. También fueron escasas la violencia y las manifestaciones de poder, pero cuando hubo violencia en general era simulada. En conclusión, estos videos latinoamericanos reflejan una sociedad en la que hay variedad de prácticas sexuales de mutuo acuerdo que no difieren de lo esperado según el modelo biológico y la antropología.

**PALABRAS CLAVE**

Videos caseros, videos aficionados, videos de internet, patrones sexuales, papeles sexuales

The subject of how human sexuality is presented in Internet has been studied for decades, and the current body of knowledge was recently reviewed by Nicola (2009) for six areas: "pornography", shops, work, education, contacts and subcultures. That review states that online activities related with sex are now routine for many in the Western world, with a frequency that depends on age, gender and sexual orientation. According to that author, many publications are biased toward a non-scientific, conservative approach, and little research has been conducted on the potential benefits of Internet sexuality. The very use of the term "pornography" rather than more rigorous terms for specific products shows a basic weakness of the academic literature on this subject that represents billions of dollars and that has extended from the old male-centered products to a large scope of products addressed to both men and women (Vera-Gamboa, 2000).

The lack of scientific protocols has led to the generalized labeling of graphic sexual materials as sexist and harmful, but some academic communities conceptualize these products in terms of a dynamic nexus of actors, discourses, media economies, technologies and consumers that can only be studied within a proper cultural frame (Paasonen, 2009).

An oversimplifying polarity is frequent in the literature: "professional and amateur", "alternative and mainstream", "independent and commercial". Feminist debates on "pornography" have been interesting but they generally could not go beyond the affective range of disgust versus pleasure (Paasonen, 2007). In conclusion, a better understanding of the subject requires less dualistic frameworks for conceptualizing Internet sexual materials as an element of media culture (Vera-Gamboa, 2000; Paasonen, 2010).

Social research on the consumption of graphic sexual products has seldom considered sexual roles, but from the few studies available, it appears that such products can generate new tastes and even subverse traditional gender roles (Figari, 2008). In sexual products, displays of control should not be automatically and simplistically translated as displays of power (Paasonen, 2006), and on the Internet the overlap between forms of sexual power is often complex and multi-directional (Brickell, 2012).

According to McKee (2009), social sciences insist that all new work agrees with previous interpretations, specially in relation to sexual subjects, for which the belief is that sexual products have negative effects. The result is an abundance of pre-Foucauldian assumptions that obscure understanding of new media and Internet sexuality. Internet sexual products must be understood considering the encoding and decoding processes and discourses of the sexuality of societies in which they occur (James & McCreadie, 2004).

We are aware of no studies about Latin American sexual videos posted in Internet and this article is a first step to fill that gap in our knowledge. We analyze the results from biological and anthropological viewpoints because we believe that a more comprehensive approach benefits our understanding of a phenomenon that has both biological and cultural aspects.

## MATERIALS AND METHODS

Nereliatube (http://thedomainfo.com/nereliatube.com/) was a site registered in Barcelona, Spain that published sexual videos from Latin America, mostly from Mexico, Venezuela and Argentina. Users who uploaded the videos were not paid but frequent contributors were offered free membership. New videos were moved from the free to the paid section after a few weeks. Unfortunately, the site closed down in 2012: our study rescues data from a site that is no longer available.

For our research we excluded all commercial videos, videos that showed the sexual activity of a single person and videos whose origin could not be identified with certainty. To identify the persons as Latin Americans we did not use the video titles because they were unreliable (we often found the same video under a variety of mutually excluding titles about who they showed and where they had been recorded). Instead we used the audio to identify nationality from the accents, an easy task for us because we are native speakers and have lived in, or visited, several parts of Latin America. We viewed all videos posted in the website and extracted 214 that fit the above criteria. The extraction was finished on December 31, 2010 and was done from the totality of videos posted in the site (more than a thousand).

We produced a correlation matrix for all possible variable interactions with the free online software VassarStats (http://vassarstats.net/). The Results section presents only variable pairs that had an statistically significant correlation of 40% or more (significance level: p<0,05).

We tested 11 hypotheses, suggested to us by our own knowledge of human sexuality in Latin America:

1. Latin American society sees a wide sexual experience in men as positive and thus men are more interested than women in recording their sexual activity, thus they are more likely to be the camera operators.
2. Women are usually afraid of social hostility if their sexual videos reach the public, so in most cases the videos will be recorded secretly by the men.
3. The videos reflect the frequency of sexual practices and thus oral sex and vaginal penetration are common.
4. Couple sex is more common than group sex because the videos reflect common practice.
5. Violence is rare because most videos are from couples in a romantic relationship.
6. Signs of affection are common (see Hypothesis 5).
7. Manifestations of power are rare (see Hypothesis 5).
8. There is more violence in couple videos than in group videos because of group pressure for a peaceful behavior.
9. Violence and demonstrations of power are correlated.
10. Violence is less frequent when women dominate numerically because women tend to be less violent than men.
11. Power demonstrations are less frequent when women dominate numerically because women use power demonstrations less markedly than men.

## RESULTS

In most cases the camera was operated by the man (N=80% of the 214 videos); this category was distantly followed by videos in which it was not clear who was operating the camera (14%), videos in which both partners operated the camera at some time (4%) and the few cases in which the woman was the operator (3%).

Sexual encounters can be recorded without the consent of some participants by using hidden cameras, but that was rare: in most videos (83%) it is clear that the camera is in plain view of all participants, and sometimes they stare at the camera or reposition it when there is a change of scene. However, in 7% of cases the camera seemed to be hidden (the man turned it on before the woman entered the scene and there was no further interaction with the camera until after she had left) and in 3% cases we could not distinguish if the camera was visible.

The videos showed a variety of normal human sexual practices. The most frequent was fellatio (56%), followed by vaginal penetration with the woman on top and facing the man (34%), with a nearly equal number of videos of vaginal penetration with the man on top (33%) and vaginal penetration with the woman on her fours and the man behind (29%). Less frequent were female masturbation (14%), female striptease (8%), male masturbation (8%), vaginal penetration with the woman on top and her back towards the man (7%), vaginal penetration with the woman lying on her side and the man behind her (5%), cunnilingus (4%) and anal penetration of the woman (3%).

Most videos showed female-male couples; others had only one case each: two women; two women and two men; five men and one woman; four men and two women and finally, seven men with one woman (Fig.1).

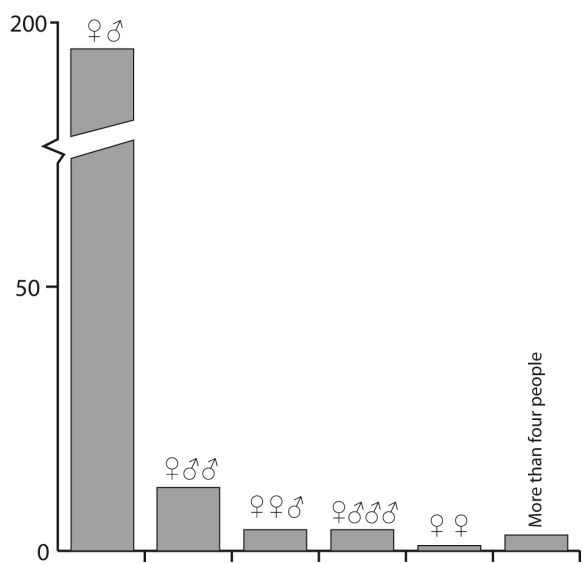

**FIG. 1.** Composition of couples and groups, and frequency of each in sexual videos.

Violence was absent in the great majority of videos (97%). There was simulated violence, in the form of slapping, spanking and pulling the woman's hair (both partners appeared to enjoy the practice), in 7% of videos. We also found two cases of real violence. In one video, a woman hit the other woman's face with her open hand and inserted a bar of soap in the victim's vagina. In an unrelated video, a drunk woman was dragged on the floor and water was poured on her face and body.

Unexpectedly from videos that seem to mainly show the sexual activity of steady couples, most did not show stroking, tender kissing or other signs of affection (97%).

There were no manifestations of power in most videos (97%); in the few that did (3%), these manifestations took the form of men forcing women to do sexual practices that they clearly refused, or using their physical strength to subdue the women.

There were four cases in which two video characteristic were correlated. There was less violence when there were more people participating in the sexual activity (Spearman Rank Correlation -0,499). When one of the partners, normally the man, made a show of his power, it was more likely that the video would also show some violence (Spearman Rank Correlation 0,425). Furthermore, groups with more women had less violence (Spearman Rank Correlation -0,425) and less power demonstrations (Spearman Rank Correlation -1,000).

## DISCUSSION

Earlier authors have not found a simple relationship between Internet sex videos and corresponding social structures, particularly power relations associated with gender or social class (Orgaz & Martínez-Novillo, 2010; Paasonen, 2010). Our results agree with the hypothesis that men are more interested than women in recording their sexual activity, probably because Latin American societies only accept promiscuity for men.

Doorn (2010) stated that, rather than providing a space for alternative sexual representations, amateur video websites are sites where sex, participatory media and the representation of reality converge to maintain a politically conservative gender ideology. We cannot say that such a result applies to the videos from Latin America that we studied. We found that women in the videos generally appear as active partners that enjoy the sexual activity being recorded, a behavior that hardly fits on the patriarchate vision of sex for reproduction with a passive woman and an active man.

The web allows the publication of sexual activities from marginalized groups and thus Internet sexuality can have more authentic representations of gender and sexuality than conventional mainstream media (van Doorn, 2010). The videos we analyzed reflect usual sexual practices of current Latin American society as we know it, they do not have the beautiful bodies and exaggerated vocal reactions of commercial videos, and thus, if overweight partners and common sex are marginalized, these videos indeed fit the statement made by van Doorn (2010). The internet is an increasingly important enabler and mediator of sexual relations in society, and it can transform earlier modes of knowing, experiencing and organizing sexuality (Brickell, 2012) Until now, most products follow a masculine, heteronormative "script" but Internet can potentially open up spaces for the sexual emancipation of groups that are normally excluded from mainstream sexual products (Orgaz et al., undated; van Doorn, 2010).

In their home-made sexual videos, women and men somehow negotiate their public presentations and gender performances (Berkowitz, 2006). Bell (1999) made an interesting comparison of weddings and graphic sex products, and found that both are cultural performances designed to solidify the traditional social organization through the control of sex; in other words, both are used to maintain the cultural status quo. Again, we found that the "performances" in Latin America videos appear to represent real life sexuality, and for that reason, violence, group sex and similar practices that are less acceptable socially were rare in the videos.

From the videos in Nereliatube we could not reject most of our hypotheses. The exceptions were hypotheses numbers 2 and 6 and these seem even more interesting because they deviate from expectations. We expected men to secretly record most videos because women would be afraid of social hostility if these reached the public. Perhaps these women believed that the videos would remain private, but we do not have rigorous data to support this or any other explanation, so the subject becomes particularly interesting and waits future research. Regarding hypothesis number 6, we do not know why signs of affection were rare in these videos, but we offer future researchers a hypothesis: that the intent was only to record the sexual encounter itself and the loving foreplay took place before the camera was turned on. Anthropologically we expected men and women to follow the socially asigned roles of active/passive with casual sex regretted more intensely by women (Galperin et al., 2012), who would also show more emotions and affectivity (Rodríguez, 2009). These trends were not strong in our results and need further research.

Earlier studies, which did not include Latin America, reported that consensual and nonconsensual violence were more frequent in user posted products than in commercial products (Barron M. & Kimmel, 2000; Dunbar & Burgoon, 2005). In our results, violence was infrequent, and when present, it was mostly consensual. We believe that these home-made videos reflect the normal behavior of Latin American couples and additional study is needed to see if the difference with Anglo-Saxon practices reflects real cultural differences or methodological differences.

The model of parental investment and sexual selection explains why men are more dominant and violent: to establish relationships with high investing females (Buss, 2006). Our results agree with this model.

In conclusion, Latin American home-made sexual videos posted in Nereliatube.com generally reflect a society in which women and men have a variety of sexual practices that are mostly consensual and that do not differ from the biologically expected patterns. Of course other authors may disagree with our hypotheses, and may even imagine additional hypotheses, but the main contribution of our study is that instead of trying to prove a particular position, we tested our hypotheses against the data and rejected some based on that test. We hope other researchers will do the same.

## ACKNOWLEDGEMENTS

We thank several anonymous reviewers for comments to an earlier draft.

*Article edited by Bernal Morera Brenes*